\documentclass[journal=jacsat,manuscript=article]{achemso}
\setkeys{acs}{etalmode=truncate,maxauthors=10}
\usepackage{chemformula} 
\usepackage[T1]{fontenc} 
\makeatletter
\def\acs@author@fnsymbol#1{} 
\makeatother

\author{Damiano Ricciarelli,\textsuperscript{\dag, \textasteriskcentered} Ioannis Deretzis,\textsuperscript{\dag} Gaetano Calogero,\textsuperscript{\dag} Giuseppe Fisicaro,\textsuperscript{\dag} Enrico Martello,\textsuperscript{\dag} Antonino {La Magna} \textsuperscript{\dag, \textasteriskcentered}}

\affiliation{\textsuperscript{\dag} National Research Council of Italy, Institute for Microelectronics and Microsystems (IMM-CNR), Z.I. VIII Strada 5, 95121 Catania, Italy}
\email{'see_IMM-CNR_website'}

\title{Local Coordination Modulates the Reflectivity of Liquefied Si-Ge Alloys}
\begin{document}

\begin{tocentry}
\includegraphics[scale=0.26]{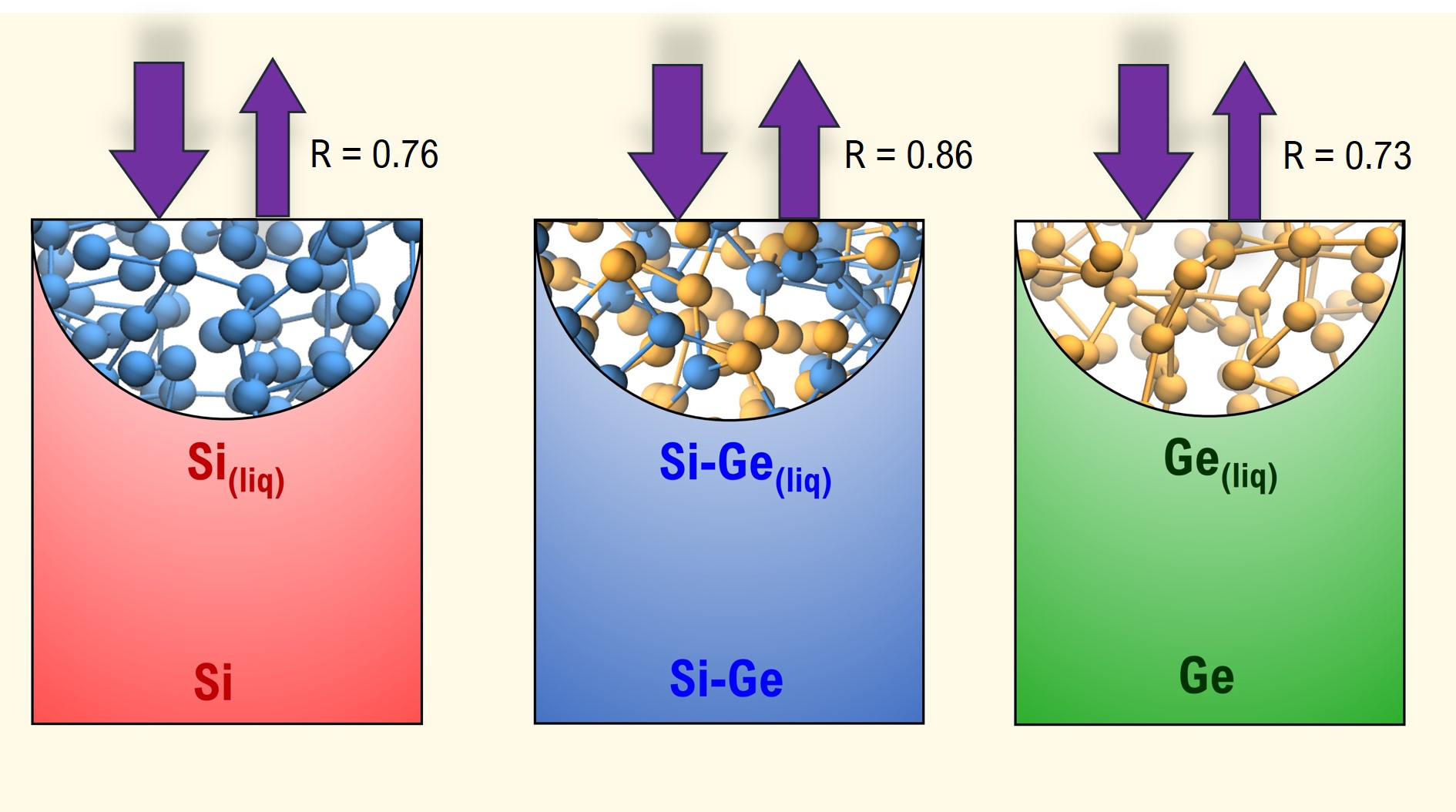}
\end{tocentry}

\begin{abstract}
The properties of liquid Si-Ge binary systems at melting conditions deviate from those expected by the ideal alloy approximation. Particularly, a non-linear dependence of the dielectric functions occurs with the reflectivity of liquid Si-Ge reaching a maximum at  50\% Ge content, being 10\% higher than in pure Si or Ge. Using \textit{ab initio} methodologies, we modelled liquefied Si-Ge alloys, unveiling very high coordination numbers and poor symmetry in the first coordination shell with respect to Si and Ge, related to different bonding properties. We simulated optical functions, quantitatively replicating the aforementioned reflectivity trend and we highlighted a direct relationship between atomic structure and optical properties, indicating that the unusual optics arises from Si-Ge higher local coordination characterized by low symmetry. We forecast further implications for the overall class of these alloys. These findings expand our comprehension of liquefied semiconductors and are essential for implementing controlled laser melting procedures to highly dope these materials for advanced transistors, superconductors, sensors and plasmonic devices.
\end{abstract}

\section{Introduction}
Group IV based materials constitute an intriguing type of systems since they manifest a variety of crystal structures, allotropes and polytypes with a consequent richness of related electroptical properties \cite{deinzer, Lee_2017, botti, sindona2022dielectric, ORTENBURGER1972653,PhysRevLett.108.067402, PhysRevLett.121.265701}. The aforementioned abundance in structures extends to the liquid states of this class of materials, however, the high temperatures needed for accurate measurements make these liquids difficult to be analyzed experimentally, and from the modeling perspective, they exhibit complex bonding regimes that are also challenging to be handled \cite{PhysRevLett.108.067402, PhysRevLett.121.265701, stich_1991, kresse_1994, ko_2002, ricci2014, chathot2009, BEAGLEHOLE1970272, PhysRevB.57.12410, PhysRevB.68.235207, PhysRevLett.76.2077, wang}. It is well known that the study of these systems has a technological relapse: Si is the mainstream material for microelectronics and photovoltaics. To diversify and enrich its properties, Ge and Si-Ge alloys are nowadays employed in many domains, enabling to modify strain, carrier mobility and optical band gaps, as Ge has a higher hole mobility, a smaller band gap and a larger lattice parameter than Si\cite{Manku1993, Jain_1991, Iyer1989, people1986, pearsall1989}.
Understanding the complete range of physical properties exhibited by the Si-Ge binary system is particularly important for technological applications. This importance not only applies to their crystalline phases but also extends to their liquid phase. Industrially, the Czochralski method is the primary process used to grow single Si crystals from a liquid state. However, the significance of liquids has greatly increased due to the utilization of ultrafast pulsed laser melting techniques in semiconductor processing, enhancing the electrical properties of the material in specific regions. The applications are diverse: initially, laser melting was used to increase the quantity of active dopants and improve the crystal quality of the material during re-growth  \cite{Baeri_1981, ong2004, hernandez2004, svensson2005, albenze, dagault2019, dagault2020_2, huet2020, Calogero_2023, calogero2022multiscale}. Recently, the non-equilibrium regime created by pulsed lasers has also been exploited to achieve conditions of ultra-doping and hyper-doping. In these cases, the concentration of active dopants can reach atomic percentages as high as 5-10\%, exceeding the crystal solubility limit. Ultra-doping involves utilizing high doping concentrations to achieve superconductivity, whereas hyper-doping is linked to augmenting semiconductor bands to increase absorption coefficients and enhance plasmonic effects in the infrared region \cite{bustarret, Grockowiak_2013, Daubriac_2021, dumas, prucnal, Buonassisi, Zianni, chao5, Poumirol}. A successful completion of  laser melting processes requires an in-depth design of experiment, often coupled to precise calculations. To this purpose, having an exhaustive comprehension of crystalline and liquid optical functions is essential. Spectroscopic ellipsometry makes it feasible to experimentally measure these values for solids \cite{Jellison_Si_1982, Vina_1984, jellison_1993}. However, for liquefied semiconductors, the process is more challenging due to the high melting temperatures needed. Over the past forty years, time-resolved spectroscopic ellipsometry has been utilized to analyze the dielectric properties of liquid Si and liquid Ge when subjected to laser-induced melting. However, absolute dielectric function values from these measurements can be derived only indirectly through extensive calibrations \cite{jellison_trr_1986, jellison_tre_1987, boneberg, vcerny1995determination, abraham}. Furthermore, there is a lack of research dedicated to determining the optical functions of liquid Si-Ge alloys. This omission is significant because the segregation of Ge between the liquid and solid phases adds a higher level of complexity to the analysis  \cite{dagault2019, dagault2020_2}. An intriguing outcome deriving from the aforementioned time-resolved spectroscopy studies on liquefied Si and Ge is the manifestation of metallic-type dielectric functions, aligning with the form predicted by Drude theory. Recently, we studied XeCl excimer laser melting with a radiation energy of 4.02 eV  on Si-Ge alloys \cite{ricciarelli2023}. On the basis of experimental data, we derived a semi-empirical formulation of liquid Si$_{\rm 1-x}$Ge$_{\rm x}$ dielectric function. This led to an unexpected trend of reflectivity (R) for liquefied Si$_{\rm 1-x}$Ge$_{\rm x}$. Specifically, at the melting temperature, Si$_{\rm 0.5}$Ge$_{\rm 0.5}$ displayed the highest reflectivity value, reaching up to $\sim$ 0.85, while Si and Ge exhibited $\sim$ 0.77 and $\sim$ 0.75 respectively. Furthermore, we found a more complex temperature dependent liquid Si$_{\rm 0.5}$Ge$_{\rm 0.5}$ reflectivity above the melting point with respect to Si and Ge. At 80-100 K above the melting tempetrature, R  Si$_{\rm 0.5}$Ge$_{\rm 0.5}$ drops to a plateau of $\sim$ 0.80 \cite{ricciarelli2023}. \\
In this work we rationalize the aforementioned reflectivity change at the melting point by means of \textit{ab initio}  molecular dynamics simulations of liquefied Si$_{\rm 1-x}$Ge$_{\rm x}$. We focused on three systems: pure Si, Si-Ge alloy and pure Ge. For simplicity, we considered 50\% Ge content in the Si-Ge alloy whom represents the intermediate between the two, where, we know, R presents a maximum. We compared their optical and structural properties, achieving a perfect agreement with the experiment \cite{ricciarelli2023} and we further underlined a clear connection between structural properties and reflectivity. \textit{Ab initio}  molecular dynamics (MD), employing either the Car-Parrinello and the Born-Oppenheimer formulations, has been widely applied to gain deeper insights into the structural, electrical, and thermal properties of the three mentioned systems \cite{stich_1991,  kresse_1994, ko_2002}. For our modeling, we chose the Car-Parrinello MD approach, making use of fairly large supercells.

\section{Results and Discussion}

We performed Car-Parrinello MD, making use of 288 atoms $3\times3 \times4$ supercells, shown in Figure~1a-c. To obtain reliable disordered structures we followed the MD methodology proposed by Stich \textit{et al.} \cite{stich_1991, CP}, consisting in rapidly randomizing the crystalline face-centered cubic configuration with very high temperatures, \textit{i.e.} 6000K, and subsequently quenching the system at the temperature of interest. From the technical side, the initial randomization covered a total time of 1.0 ps and was followed by 1.5 ps of equilibration and by 3.5 ps of production run (for further details see Figure S2, Supporting Information). Then, we determined the dielectric function of those liquids, employing the random-phase approximation (RPA) approach,  making use of random representative snapshots from the MD\cite{brener_1975}. Complete information on the entire computational details that involves the Perdew-Burke-Ernzerhof functional and D3 Van der Waals corrections within the Quantum Espresso software can be found in the Supporting Information \cite{pbe_1996, vdb_us, grimme, vanderbilt_1990, QE}.\\
We employed a canonical fixed-volume NVT ensemble setting equilibrium temperatures as the melting points, namely 1687K (Si), 1543K (Si-Ge), and 1211K (Ge). We fixed cell parameters, as shown in Figure~1a-c, to match liquids' densities reported by Ricci \textit{et al.} \cite{ricci2014}, corresponding to 2.5 (Si), 5.9 (Si-Ge) and 5.6 (Ge) $\rm g \cdot \rm cm^{-3}$.  
Comparing these densities to crystal's ones \cite{wyckoff}, we observed remarkable differences in the melting volume shrinkage for the three liquefied semiconductors. Si (Ge) shows only 7\% (5\%) contraction, whereas Si-Ge experiences a higher contraction of 35\% (for additional details see Figure~S3, Supporting Information). \\
\begin{figure*}
\includegraphics[scale=0.45]{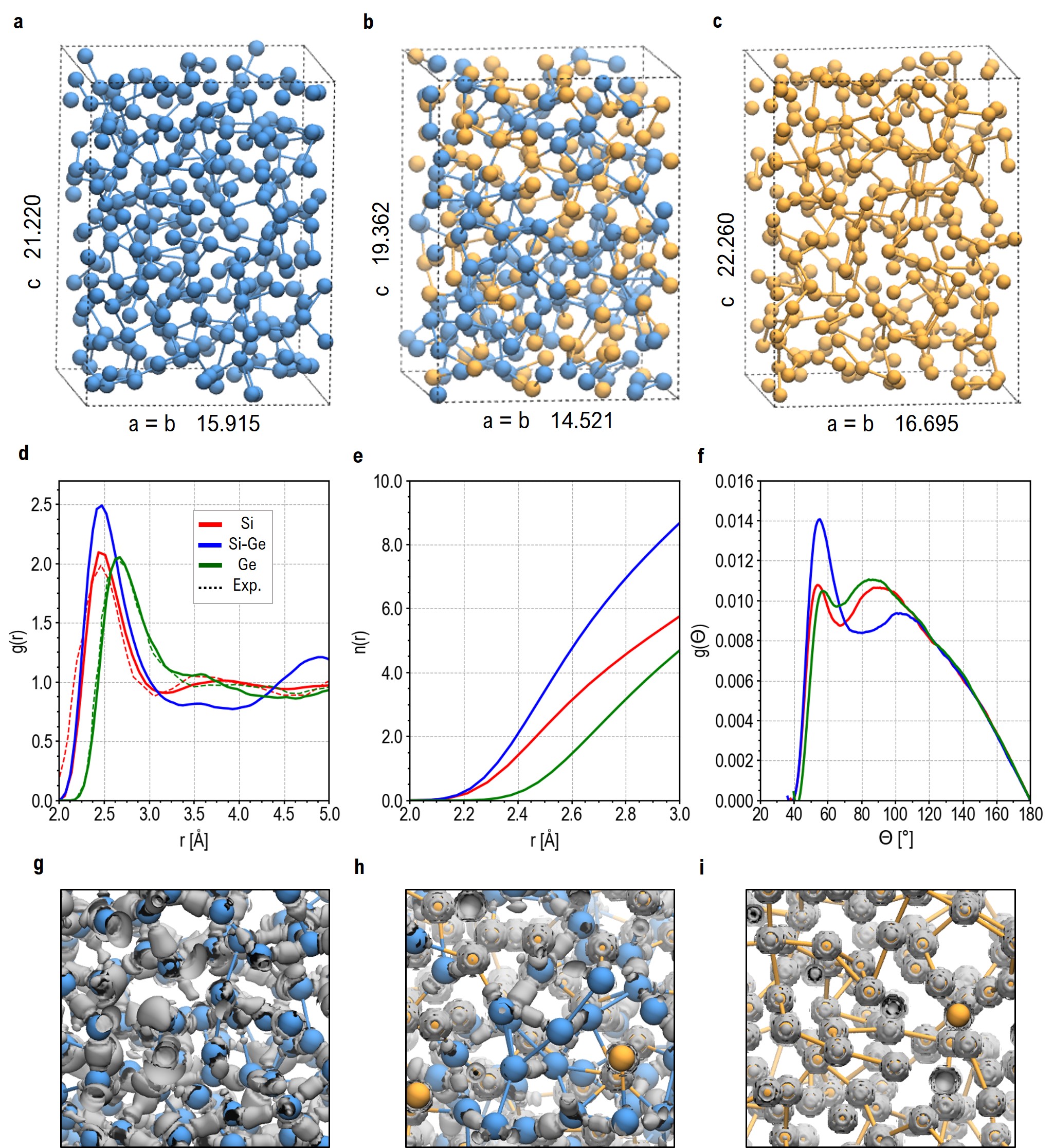}
\caption{Supercells employed to model (a) liquid Si, (b) Si-Ge and (c) Ge. Key functions for the structural analysis of liquids computed at their melting temperature: (d) radial distribution function, $g(r)$,  (e) coordination number, $n(r)$, and (f) angular distribution function, $g (\theta)$, for $R_{cut}$$<$3 \AA   . Solid lines in $g(r)$ plots are associated to our simulations, while dashed lines to neutron diffraction experiments \cite{Gabathuler, Salmon_1988}. Iso-surface plots of electron localization function (iso-value = 0.78) for (g) liquid Si, (h) Si-Ge and (i) Ge at their melting temperature computed on selected snapshots. For more info \textit{cf.} main text.}
\end{figure*}
We utilized key functions to examine structural properties in our MD study: the radial distribution function $g(r)$, representing the likelihood of finding a particle at a distance \textit{r} from a reference particle, $n(r)$, indicating the number of coordinated atoms within a sphere of radius \textit{r}, and the angular distribution function $g(\theta)$, determining the probability of bond angles \cite{chandler1987introduction, stich_1991}. The  $g(r)$ functions of liquid Si and Ge, shown in Figure~1d, closely matched the first coordination shell peak measured by neutron diffraction \cite{Gabathuler, Salmon_1988}, while, to the best of our knowledge, the experimental $g(r)$ of Si-Ge has never been reported. Some uncertainty occurred for the reproduction of Si and Ge second coordination shell peak, we are unable to disentangle whether this comes from the finite size of the liquid's cell employed or from errors associated to the experimental measurement. Nevertheless, the first peak, which we successfully replicated, is where the gross of bulk liquid's properties originate.  From our simulations, the first Si (Ge) coordination shell peak was found at $\sim$ 2.5 Å, ($\sim$ 2.7 Å), and the small variation was attributed to stronger bonds in Si, similar to the crystals' picture \cite{Buriak}. In the Si-Ge binary system, the first $g(r)$ peak was aligned with Si rather than falling at an intermediate distance, signaling a Si-Ge bond energy comparable to Si. For both the first and second coordination shells, the intensity of the Si-Ge $g(r)$ function was $\sim$ 25 \% higher than that of pure elements. This heightened intensity reflected an enhanced local coordination, which was also observable in terms of the $n(r)$ functions, Figure~1e, following the trend $n(r)$ Ge $<$ $n(r)$ Si $<$ $n(r)$ Si-Ge. From computed coordination numbers (CN), we observed striking differences: Si (Ge) exhibited a CN of 5.7 (4.8), consistent with previous research \cite{stich_1991, kresse_1994}, while, Si-Ge showed a remarkably higher value of 8.6. Furthermore, as depicted in Figure~1f, the prominent $g(\theta)$ peak experienced by the alloy at $\sim$ 50° assessed, in accordance with the guideline established by Ishimaru \textit{et al.} \cite{ishimaru}, that the system presents a high coordination, with atoms occupying very narrow positions relatively to each other. The $g(\theta)$ function further indicates that the distribution of bond angle in Si-Ge is uneven, suggesting a lack of symmetry within the first coordination shell.\\
On overall, these data demonstrate that Si-Ge exhibits higher coordination than pure elements due to the formation of a closer packed first coordination shell, whom indeed stems from the chemical bonds involved in Si-Ge. By decomposing the Si-Ge $n(r)$ function into inter- and intra-element contributions in Figure~2 we found the first coordination shell to be evenly split between Si-Si (or Ge-Ge) and Si-Ge bonds, implying that the liquid alloy exhibits an equal number of bonds between similar and dissimilar atoms at the same time, almost twice as many as those experienced by the two liquid elements.

\begin{figure*}
\includegraphics[scale=0.50]{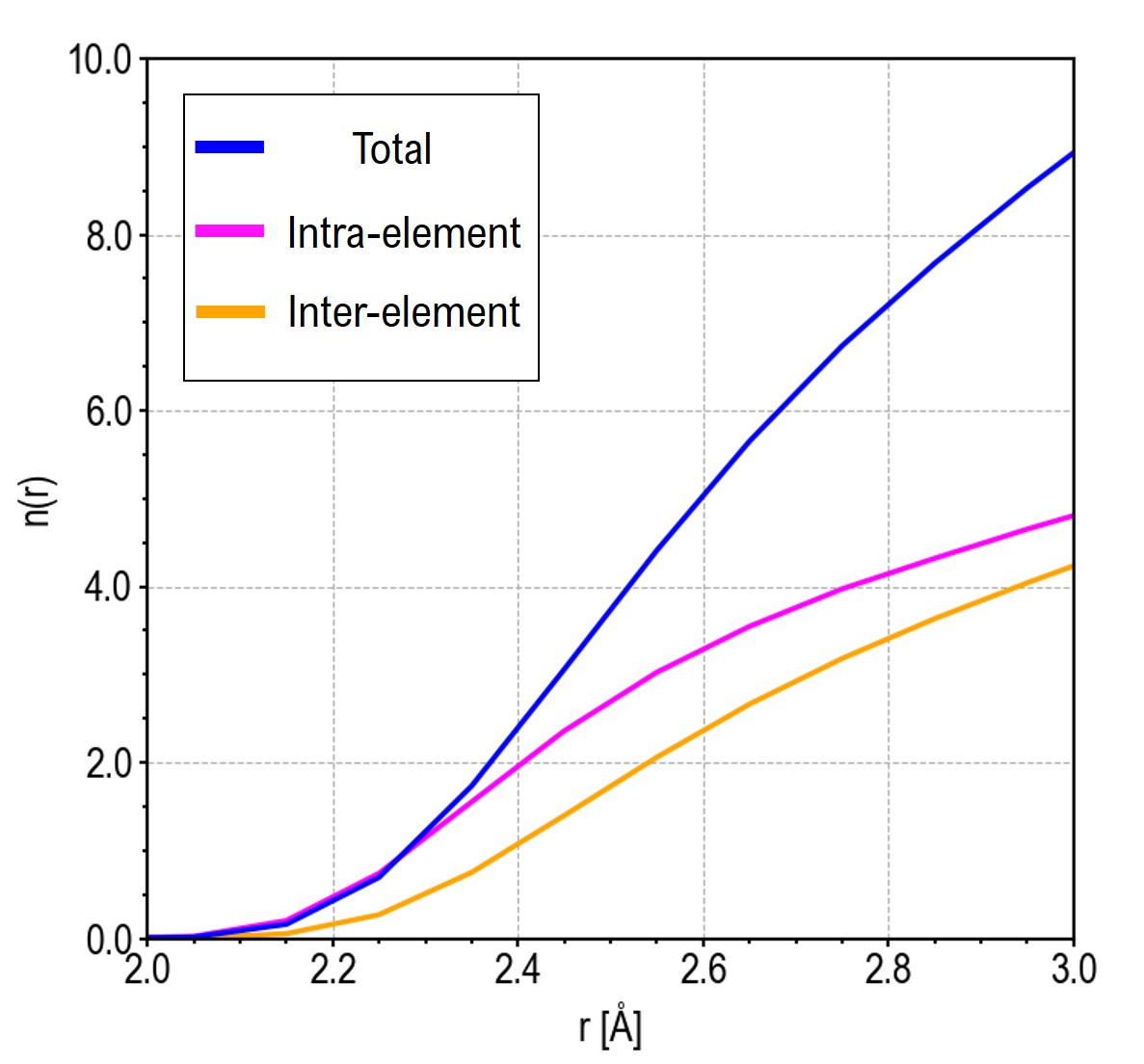}
\caption{Decomposition of the $n(r)$ function of Si-Ge into intra-element and inter-element contributions.For more info \textit{cf.} main text.}
\end{figure*}

In order to gain more information on the chemical bonds type, we plotted electron localization functions (ELF) in Figure~1g-i \cite{savin}. Si and Ge displayed distinct bonding patterns: Si exhibited more defined and directional $\sigma$-type bonds, reflecting higher Si \textit{s} and \textit{p} orbital hybridization, while, Ge had less defined bonds with localized electron density on non-hybridized Ge \textit{s} orbitals. Si-Ge displayed intermediate features, showcasing both $\sigma$ and \textit{s} patterns (Figure~1g-i) but ELF planar averaged profiles in Figure~S4, Supporting Information, signaled an electron density distribution similar to Si, with mean values of $\sim$ 0.40. Summing up all these data, the increased coordination in the alloy stem from Si atoms in the first coordination shell prioritizing the formation of stronger, more directional \textit{$\sigma$} bonds with neighboring Si atoms, while, also bonding with Ge atoms. This would nearly double the total number of bonds, being at the origin of the overall enhanced Si-Ge coordination.  \\
The liquid alloy structural properties differ from those of crystals, where bond lengths follows the trend Si (2.31 Å). $<$ Si-Ge (2.42 Å) $<$ Ge (2.50 Å), CNs are $\sim$ 4.0 for both Si-Ge and pure elements and bonding energies/patterns change linearly with the increased Ge content (For more details see Figure~S5, Supporting Information).\cite{wyckoff} \\
It is worth noting that our Si-Ge structural results are slightly at variance with the ones of Ko \textit{et al.}, where the supercell volume was calculated as an average between the pure elements, without considering the specific experimental density \cite{ko_2002}. Despite the first $g(r)$ peak was detected at 2.5 \AA, the computed coordination number of $\sim 6.5$ was significantly smaller than ours. \\
\begin{figure}
\includegraphics[scale=0.50]{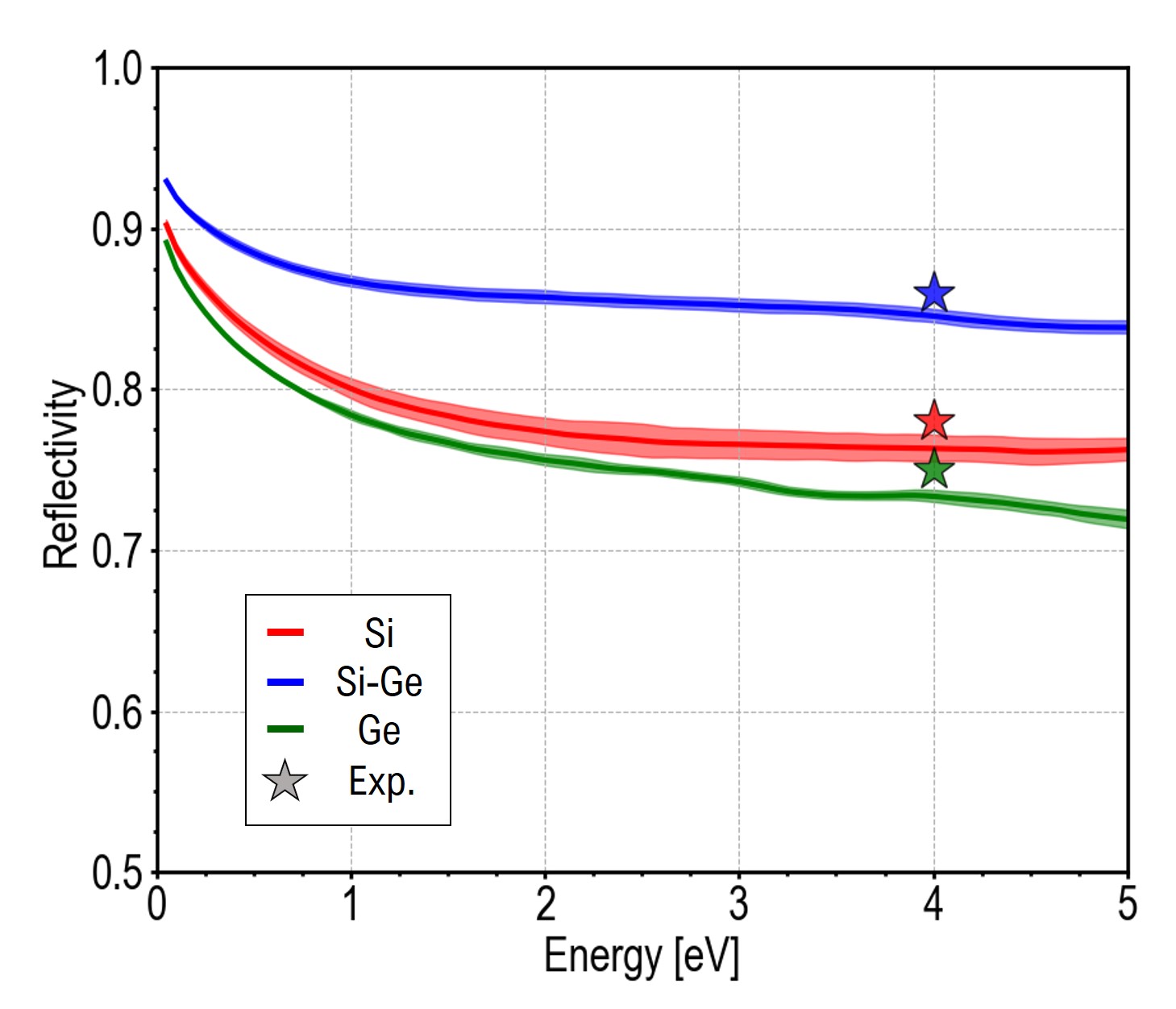}
\caption{Reflectivity vs excitation energy functions of liquefied Si, Si-Ge and Ge at their melting temperature. Filled area and solid lines represents the overall statistic intervals and average values respectively from our study, while stars represent R values taken from the liquid reflectivity map drawn by Ricciarelli \textit{et al} \cite{ricciarelli2023}. For more info \textit{cf.} main text.}
\end{figure}
Having understood the structural and bonding properties of the liquids at their melting point, we turned to optical functions. The reflectivity (R) trends at the melting point, shown in Figure~3, presented a fairly good agreement with results from our previous experiment-based study \cite{ricciarelli2023}, in which a radiation energy of 4.02 eV was used, yielding  a maximum R value of 0.86 for Si-Ge and 0.76 (0.73) for elemental Si (Ge). Importantly, in the overall range of considered excitation energies, R Si-Ge exceeded R Si and R Ge, whom in turns show similar values, following $R_{Si-Ge} > R_{Si}, R_{Ge}$ reinforcing the correctness of the trend gained previously. \cite{ricciarelli2023} It is worth to highlight a subtle variation between Si and Ge due to a reduced plasmon energy. As shown in Figure~S6, Supporting Information, the reflectivity behavior of the three liquefied semiconductors is substantially different from that of crystalline systems, whose R functions are overlapping, averaging $\sim$ 0.50. \\
To deepen our understanding of optical results, we analyzed the real and imaginary components of the dielectric functions in Figure~4a-b. We employed a Drude model to fit the real part of dielectric functions, this model is represented by equations (1)-(2),

\begin{equation}
\Re (\varepsilon)=1-\omega_p^2 \cdot \frac{1}{\omega^2+ \Gamma^2}
\end{equation}
\begin{equation}
\Im (\varepsilon) = \omega_p^2 \cdot \frac{\Gamma}{\omega(\omega^2+\Gamma^2)}
\end{equation}

where $\omega$ is the excitation energy, $\omega_p$ is the plasmon energy, and $\Gamma$ is the damping factor.

\begin{figure*}
\includegraphics[scale=0.45]{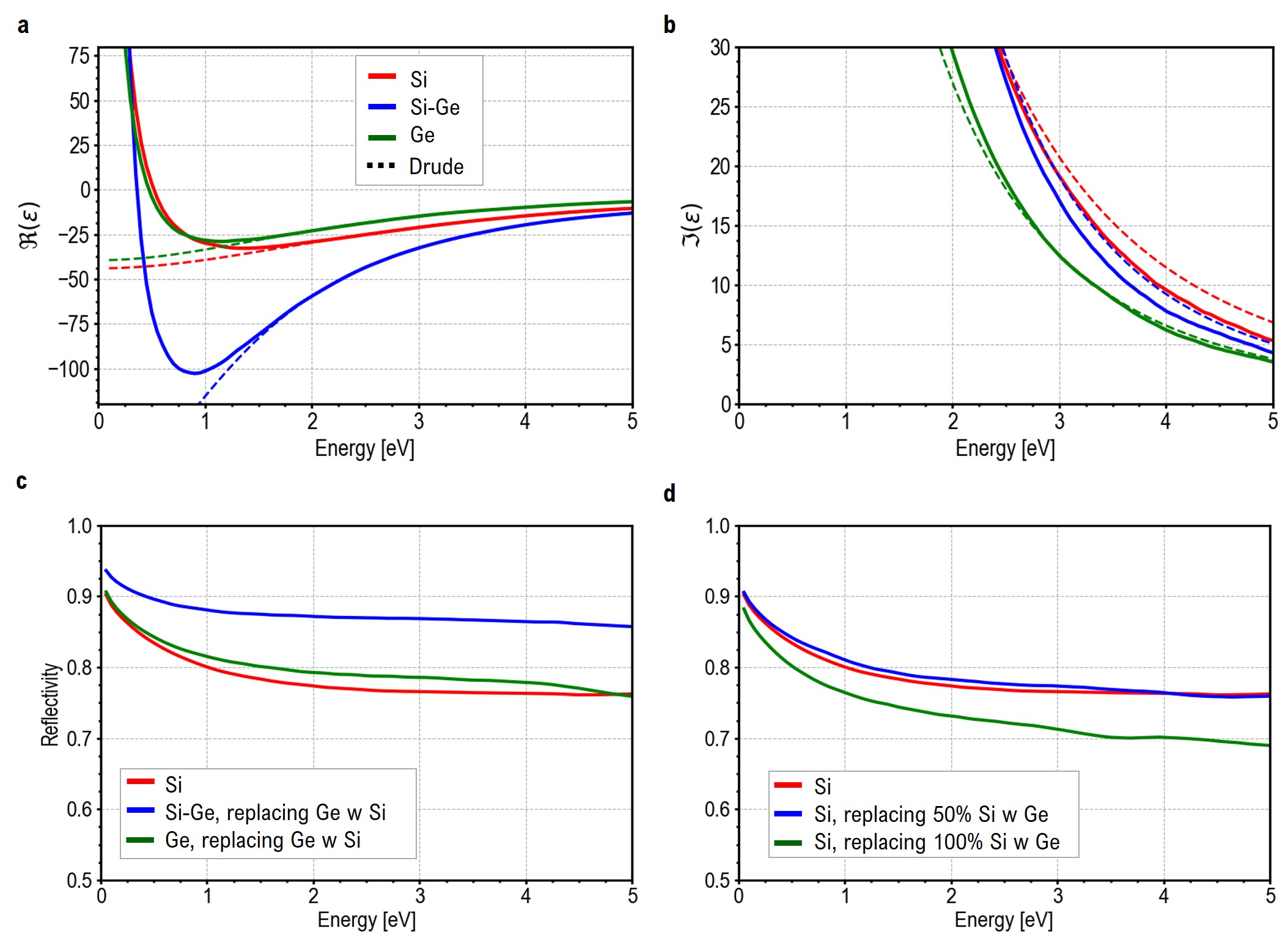}
\caption{(a) Real and (b) imaginary parts of dielectric function at the melting point, solid lines represents values from our simulations, dashed curves are obtained by fitting $\Re (\varepsilon)$ with the Drude model. Analysis of reflectivity trend: (c) averaged R computed employing liquid structures, substituting Ge atoms with Si to isolate the impact of structures on R, (d) averaged R computed employing Si liquid structures, substituting Si with Ge atoms to match Si, Si-Ge and Ge stoichiometry (coordinates and cells are further re-scaled to account for the cell parameters' differences) to isolate composition's effects on R. For more info \textit{cf.} main text.}
\end{figure*}

Consistently with previous studies \cite{jellison_tre_1987, jellison_trr_1986}, the Drude model accurately described the optical behavior of these molten alloys, \textit{i.e.} they manifest a metallic character. Specifically, this occurred for excitation energies exceeding 1.5 eV. Table~1 provides the values of $\omega_p$, $\Gamma$, and $\tau$ (average electron collision time). By analyzing the real part of the dielectric function in Figure~4a, we observed a striking similarity between Si and Ge, reflected by similar $\Gamma$ values of 2.9 eV and 2.4 eV respectively. This similarity highlighted a comparable damping effect on electromagnetic radiation for both elements. In contrast, the real part of the dielectric function of Si-Ge displayed lower values and a smaller $\Gamma$ of 1.5 eV. This consistently describes a scenario of less damped electromagnetic radiation, indicating the presence of more forbidden intra-band electronic transitions and longer-living excited states. The notion of increased forbidden excitations in liquid Si-Ge is in line with the single-particle analysis presented in Figure~S7, Supporting Information. Virtual Si-Ge single-particle states appeared significantly out-of-phase when compared to the higher energy occupied state, reinforcing the observed trends. As mentioned, the slight differences in R values for Ge, when compared to Si, can be rationalized in terms of distinct plasmon energies. Specifically, identical values of 19.6 eV were observed for Si and Si-Ge, while Ge exhibited a slightly lower value of 15.3 eV. We noticed that the imaginary part of the dielectric function in Figure~4b is consistent with the Drude model derived from the real part.  \\

\begin{table}
\caption{Values of plasmon energy, $\omega_p$, damping factor, $\Gamma$,and averaged electron collision time, $\tau$, corresponding to $1/\Gamma$. The parameters are determined by fitting $\Re (\varepsilon)$ with the Drude model.}
\begin{center}
\begin{tabular}{ cccc } 
\hline
 Liquid & $\omega_p$  & $\Gamma$ & $\tau$ \\
 & [eV]& [eV]&[fs]\\
 \hline
 Si - 1683 K& 19.6 $\pm$ 0.2  &2.9 $\pm$ 0.1 & 1.4\\ 
 Si-Ge - 1543 K& 19.6 $\pm$ 0.1 & 1.5 $\pm$ 0.2 & 2.8\\ 
  Ge - 1211 K& 15.3 $\pm$ 0.2 & 2.4 $\pm$ 0.1 & 1.7\\
 \hline
\end{tabular}
\end{center}
\end{table}
We conducted further tests to distinguish between the influences of structural and electronic factors on the reflectivity patterns in Figure~3, to ultimately gain correlation between atomistic properties and optics. To isolate the structural effects, we computed the R of Si, Si-Ge, and Ge by substituting Ge atoms (when present) with Si in randomly collected geometries. The results (Figure~4c), replicated the trend $R_{Si-Ge} > R_{Si},R_{Ge}$. In contrast, starting with random liquid Si snapshots and replacing some Si atoms to achieve Si-Ge and Ge stoichiometry, along with rescaling the cell parameters and related coordinates, we did not reproduce the trend (Figure~4d). From these tests, we can deduce that the reflectivity behavior of the different liquids is primarily determined from their individual geometries rather than their composition. As mentioned in the initial paragraphs of the section, the structure of Si-Ge differed from the individual elements for the coordination character, specifically Si-Ge exhibits a highly coordinated shell with lower symmetry. Therefore, we can assert that, within silicon-germanium alloys, it is the coordination shell character whom governs the liquid's reflectivity: higher coordination regimes, with less centrosymmetric topologies, such as the one experienced in Si-Ge at the melting point, deliver a higher reflectivity. \\
As indicated by our previous study,\cite{ricciarelli2023} the liquid reflectivity reaches its peak at an intermediate Ge content. Building upon this knowledge, our current findings reveal a clear association between the reflectivity (R) and the local coordination. From this, we can forecast that, at the melting point, the coordination of the liquid alloy will experience an increase from 0\% to 50\% Ge content, followed by a decrease between 50\% and 100\% Ge content. This will be indeed associated to a shrinkage and an expansion of the liquid volume respectively.

\section{Conclusions}

In conclusion we compared the structural and optical properties at the melting point for liquid Si, Ge and Si-Ge from \textit{ab initio} molecular dynamics. From the structural side, our results highlight a significant difference among the three systems with Si-Ge presenting, a higher coordination number of 8.6 than 5.7 (4.8) of Si (Ge) and a lower symmetry of the shell. Concerning the optical properties, Si-Ge shows higher reflectivity values of the liquid near the melting point if compared to the pure elements, in agreement with experimental studies \cite{ricciarelli2023}. We found that this trend originates primarily from the different coordination character of the liquid rather than the different element composition. The higher coordination regime met in Si-Ge with lower centrosymmetric character provides higher reflectivity values. We found that the observed structural and optical properties obtained for liquids at the melting point are different if compared to analogous crystalline systems. Given that reflectivity reaches its maximum at an intermediate Ge content at the melting temperature\cite{ricciarelli2023} and our current findings establish that reflectivity is influenced by local coordination, we can predict a coordination augmentation in the liquid alloy from 0\% to 50\% Ge content, succeeded by a reduction between 50\% and 100\% Ge content. This will correspondingly result in a contraction and expansion of the liquid volume respectively. 
Our findings brilliantly enhance the understanding of liquefied semiconductors, paving the way to more controlled laser melting processes on Si-Ge alloys to achieve very high doping concentrations.\\
\\

\begin{acknowledgement}
We gratefully acknowledge funding from the European Union’s Horizon 2020 Research and Innovation programme under grant agreement No. 871813 MUNDFAB and and the European Union’s NextGenerationEU under grant agreement CN00000013-National Centre for HPC, Big Data, and Quantum Computing
for computational support.
\end{acknowledgement}

\begin{suppinfo}
Extended version of computational details, supplemental structural data and equation to compute $g(r)$, $n(r)$ and $g(\theta)$, supplemental electronic structure data, supplemental optical data, projected density of states and BSE benchmarks.
\end{suppinfo}

\bibliography{main}

\end{document}